\documentclass{elsart}

\usepackage{natbib}
\usepackage{graphicx}

\usepackage{amssymb}

\begin{document}

\begin{frontmatter}



\title{STM tunneling through a quantum wire with a side-attached impurity}


\author{T. Kwapi\'{n}ski, M. Krawiec, M. Ja\l ochowski}

\address{Institute of Physics and Nanotechnology Center, 
         M. Curie-Sk\l odowska University, Pl. M. Curie-Sk\l odowskiej 1, 
	 20-031 Lublin, Poland}

\begin{abstract}
The STM tunneling through a quantum wire (QW) with a side-attached impurity 
(atom, island) is investigated using a tight-binding model and the 
nonequilibrium Keldysh Green function method. The impurity can be coupled to 
one or more QW atoms. The presence of the impurity strongly modifies the local 
density of states of the wire atoms, thus influences the STM tunneling through
all the wire atoms. The transport properties of the impurity itself are also 
investigated mainly as a function of the wire length and the way it is coupled 
to the wire. It is shown that the properties of the impurity itself and the way
it is coupled to the wire strongly influence the STM tunneling which is 
reflected in the density of states and differential conductance.
\end{abstract}


\end{frontmatter}

%%%%%%%%%%%%%%%%%%%%%%%%%%%%%%%%%%%%%%%%%%%%%%%%%%%%%%%%%%%%%%%%%%%%%%%%%%%%%%%
\section{Introduction}
\label{intro}
%%%%%%%%%%%%%%%%%%%%%%%%%%%%%%%%%%%%%%%%%%%%%%%%%%%%%%%%%%%%%%%%%%%%%%%%%%%%%%%

The invention of scanning tunneling microscopy (STM) \cite{Binnig} was a 
milestone in experimental surface physics. Moreover, it became possible to 
tailor and analyze small nanostructures on various conducting surfaces 
\cite{Briggs,Hofer}. Perhaps the most spectacular and pioneering examples are 
the quantum corral experiments, in which closed atomic structures were 
assembled with help of atomic manipulations \cite{Crommie}. As the STM is a 
real space technique, and is very sensitive to the local atomic and electronic 
structures, it allows to study the properties of various, not necessarily 
periodic, structures with atomic resolution. Those include various surface
reconstructions \cite{Hofer} and low dimensional structures, like single 
adatoms \cite{Briggs,Hofer}, islands \cite{Jalochowski_1,MK_1} or 
one-dimensional monoatomic chains \cite{Himpsel}-\cite{MK_2}. 

In particular, the one-dimensional structures are very interesting from a 
scientific point of view, as they exhibit extremely rich phenomena, very often
different from those in two and three dimensions \cite{Luttinger}. However, in
reality all the one-dimensional chains always stay in contact with their 
neighborhood (substrate, external electrodes, etc.), thus usually preventing 
the observations of the exotic physics. Moreover, very often they contain 
various imperfections, like impurities, dislocations or lacks of atoms. Such a
situation likely takes place when those monoatomic chains are fabricated in
self-assembly processes. The typical examples are all one-dimensional 
structures on vicinal Si surfaces \cite{Himpsel,Crain,MK_2}. 

The electron transport through a chain in two terminal geometry, in which the 
end atoms of the chain were coupled to external electrodes has been extensively 
studied, both experimentally and theoretically (see Ref. \cite{Agrait} for a 
review). A number of experiments has revealed many interesting phenomena, like 
conductance quantization in units of $G_0 = 2 e^2/h$ \cite{Wees}, deviations 
from that ($0.7$ ($G_0$) anomaly) \cite{Thomas}, spin-charge separation 
(Luttinger liquid) \cite{Auslaender,Auslaender_1}, oscillations of the 
conductance as a function of the chain length \cite{Muller,Smit} or spontaneous 
spin polarization \cite{Thomas,Kane}. The problem of impurities or disorder in 
one-dimensional wires has also been studied both experimentally 
\cite{Thijssen} and theoretically \cite{Luther}-\cite{Fogler}. The theoretical 
studies revealed that even a single impurity can lead to a dramatic 
modifications of the low energy physics. In particular, the conductance of a 
wire with interacting impurity shows a power law behavior with the scaling 
exponent depending on the strength of the impurity \cite{Enss}.

The problem of impurities (single atoms or clusters of atoms), but coupled 
sideways to a wire, has also been extensively studied recently 
\cite{Kang}-\cite{Thygesen}. For a single impurity with a strong Coulomb 
interaction many authors predicted a suppression of the wire conductance due 
to the Fano interference between ballistic (wire) channel and the impurity 
channel \cite{Kang}-\cite{Stefanski}. To be more precise, the Fano interference 
with the impurity reverses the gate voltage dependence of the conductance 
compared to the case when the impurity is embedded in a wire. The Fano effect 
in this case can have a 'classical' nature, if the interference comes from the
impurity single particle channel, or a 'many-body' nature, when the resonant
channel is formed by the Kondo effect at the impurity. The Fano effect in a 
wire with a side coupled quantum dot has been recently confirmed experimentally
\cite{Kobayashi}.

It is the purpose of the present paper to see what modifications of the 
transport properties of a chain introduce a side coupled impurity. Here, 
however, we shall study transport properties in different geometry, in which 
all the atoms in the chain as well as the impurity are coupled to one electrode 
and the second electrode is attached to one particular chain atom or the 
impurity via additional probing atom. This geometry simply corresponds to STM 
tunneling through monoatomic chain with a side attached impurity, and can be 
related to previously mentioned STM study of self-assembled monoatomic chains 
on vicinal surfaces \cite{Himpsel,Crain,MK_2}. To our knowledge, the problem in 
this geometry has not been studied experimentally so far, except the effects of 
impurities on the length distribution of atomic chains \cite{Crain_1}.
Our system is described by tight binding model with no electron correlations, 
as we are not interested in many body effects like Kondo effect,
metal-insulator transition, spin-charge separation, etc. So our model can be
applied to the wires where the interaction energy is smaller than the kinetic
energy associated with the hopping along the wire. For example, this could 
describe situation on various vicinal surfaces (like Si(335), Si(557), etc.) 
with monoatomic Au chains grown on them, where the correlation effects are
negligible due to small carrier concentration \cite{Crain_2}. Moreover, the 
problem can be solved exactly in this case.
In order to calculate the tunneling current we have adapted a non-equilibrium 
Keldysh Green's function technique. The organization of the rest of the paper 
is as follows. In Sec. \ref{model} we introduce our model, and the results of 
the calculations and discussion are presented in Sec. \ref{results}. We end up 
with summary and conclusions.

%%%%%%%%%%%%%%%%%%%%%%%%%%%%%%%%%%%%%%%%%%%%%%%%%%%%%%%%%%%%%%%%%%%%%%%%%%%%%%

\section{The model}
\label{model}
%%%%%%%%%%%%%%%%%%%%%%%%%%%%%%%%%%%%%%%%%%%%%%%%%%%%%%%%%%%%%%%%%%%%%%%%%%%%%%%

\begin{figure}[h]
 \begin{center}
  \resizebox{0.7\linewidth}{!}{
   \includegraphics{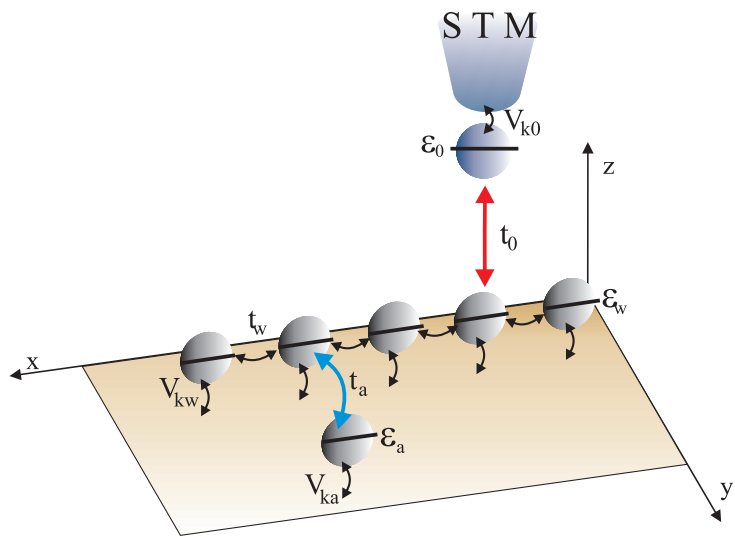}}
 \end{center}
 \caption{\label{Fig1} \footnotesize Schematic view of the model STM system 
          containing a chain with the side-coupled impurity.}
\end{figure}

Our model system is shown schematically in Fig. \ref{Fig1} and is composed of a 
wire ($w$) with atomic energies $\varepsilon_w$ and hopping integrals $t_w$ 
between nearest neighbor atoms, impurity ($a$) with single energy level 
$\varepsilon_a$, which is side-coupled to the wire via hopping $t_a$. 
Both the wire and the impurity interact with the surface ($s$) via 
$V_{{\bf k} w}$ and $V_{{\bf k} a}$, respectively. The surface is treated as a 
reservoir for electrons with the wave vector ${\bf k}$, the spin $\sigma$ and 
single particle energies $\epsilon_{s {\bf k}}$. Above the wire there is a STM 
tip ($0$) modeled by a single atom with the energy level $\varepsilon_0$ 
attached to another reservoir ($t$) (with electron energies 
$\epsilon_{t {\bf k}}$) via parameter $V_{{\bf k} 0}$. Tunneling of the 
electrons between STM tip and one of the atoms in a wire is described by the 
parameter $t_0$. The whole system is described by the following model 
Hamiltonian
\begin{eqnarray}
H = \sum_{\lambda \in \{t,s\} {\bf k} \sigma} \epsilon_{\lambda {\bf k}}
    c^+_{\lambda {\bf k} \sigma} c_{\lambda {\bf k} \sigma} + 
    \sum_{\sigma} \varepsilon_0 c^+_{0 \sigma} c_{0 \sigma} +
    \sum_{i\sigma} \varepsilon_w c^+_{i \sigma} c_{i \sigma} +
    \sum_{\sigma} \varepsilon_a c^+_{a \sigma} c_{a \sigma} \nonumber \\
  + \sum_{\sigma} (t_0 c^+_{0 \sigma} c_{i \sigma} + H.c.) +
    \sum_{ij\sigma} (t_w c^+_{i\sigma} c_{j\sigma} + H.c.) +
    \sum_{\sigma} (t_a c^+_{a \sigma} c_{j \sigma} + H.c.) \nonumber \\
  + \sum_{{\bf k} \sigma} (V_{{\bf k} 0} c^+_{t {\bf k} \sigma} c_{0 \sigma} + 
                           H.c.) +
    \sum_{{\bf k} i\sigma} (V_{{\bf k} w} c^+_{s {\bf k} \sigma} c_{i \sigma} +
                            H.c.) + 
    \sum_{{\bf k} \sigma} (V_{{\bf k} a} c^+_{s {\bf k} \sigma} c_{a \sigma} + 
                           H.c.) \; ,
\label{Hamilt}
\end{eqnarray}
where, as usually $c^+_{\lambda}$ ($c_{\lambda}$) stands for the creation
(annihilation) electron operator in the STM lead ($\lambda = t$), tip atom 
($\lambda = 0$), wire ($\lambda = i$), impurity ($\lambda = a$) and the surface
($\lambda = s$). 

In order to calculate the tunneling current from the STM electrode to the
surface we follow the standard derivations \cite{Haug,MK_3} and get
\begin{eqnarray}
I = \frac{e}{\hbar} \int \frac{d\omega}{2\pi} T(\omega) 
[f(\omega + eV/2) - f(\omega - eV/2)] \; ,
\label{current}
\end{eqnarray}
where $f(\omega)$ is the Fermi distribution function, 
$eV = \mu_t - \mu_s$ ($\mu_s = - \mu_t$) is the bias voltage, i.e. the 
difference between chemical potentials in the STM ($\mu_t$) and the surface 
($\mu_s$) reservoirs, and the transmittance $T(\omega)$ is given in the form
\begin{eqnarray} 
T(\omega) = \sum_{\sigma} \Gamma_t(\omega) \Gamma_s(\omega) 
| \sum_i G^r_{0i\sigma}(\omega) + G^r_{0a\sigma}(\omega) |^2 \; ,
\label{transmit}
\end{eqnarray}
with the coupling parameter 
$\Gamma_{t(s)}(\omega) = 2 \pi \sum_{{\bf k} \in s(t)} 
|V_{{\bf k} 0(w)}|^2 \delta(\omega - \epsilon_{t(s){\bf k}})$ between the STM
electrode ($t$) and the tip atom ($0$) and between the surface ($s$) and the
wire ($w$). Note that, to get the above expression for the transmittance we 
have assumed the same values of the coupling of the wire and the impurity with 
the surface, i.e. $V_{{\bf k} w}$ = $V_{{\bf k} i}$. $G^r_{0i\sigma}(\omega)$ 
($G^r_{0a\sigma}(\omega)$) is the Fourier transform of the retarded Green's 
function (GF) $G^r_{0i(a)\sigma}(t) = 
i \theta \langle [c_{0\sigma} , c^+_{i(a)\sigma} ]_+ \rangle$, i.e. the matrix 
element (connecting the tip atom $0$ with $i$th atom in the chain or with the 
impurity $a$) of full GF, obtained from the solution of the equation 
\begin{eqnarray}
(\omega \hat 1 - \hat H) \hat G^r(\omega) = \hat 1
\label{GF}
\end{eqnarray}
The full GF $\hat G(\omega)$ is a $(N+2) \times (N+2)$ matrix ($N$ atoms in a
chain, the impurity and the tip atom), which is obtained by inverting the 
matrix $(\omega \hat 1 - \hat H)$.

%%%%%%%%%%%%%%%%%%%%%%%%%%%%%%%%%%%%%%%%%%%%%%%%%%%%%%%%%%%%%%%%%%%%%%%%%%%%%%

\section{Results and discussion}
\label{results}
%%%%%%%%%%%%%%%%%%%%%%%%%%%%%%%%%%%%%%%%%%%%%%%%%%%%%%%%%%%%%%%%%%%%%%%%%%%%%%%

Before the presentation of the numerical results, it is worthwhile to comment 
on choice of the model parameters used in the present work. In numerical 
calculations we have assumed equal and energy independent coupling parameters 
($\Gamma_{t(s)}(\omega) = \Gamma_{t(s)}$), which reflects constant energy bands 
in both electrodes, and chosen $\Gamma_t = \Gamma_s = \Gamma$ as an energy 
unit. The other parameters have been chosen in order to satisfy the realistic 
situation in many experiments. The hopping integral within a wire is 
$t_w = 2$, $t_a = 1$, $t_0 = 0.1$, and $\varepsilon_w = 0$. For example, taking 
$\Gamma = 0.05$ eV, we get $t_w = 0.1$ eV, $t_a = 0.05$ eV, and $t_0 = 0.005$ 
eV. Such a value of the parameter $t_0$ together with the typical value of the 
work function $W = 5$ eV, gives a tip-surface distance $z = 6$ \AA 
\cite{MK_2,Calev}, and the STM current stays in the range of a few nA. These 
are typical conditions in real STM experiments. Note that, with such a value of 
$t_0$, the modifications of the wire density of states due to the STM tip are 
negligible. In the following we will discuss the properties of wires containing 
even and odd number of atoms, as we expect different behavior for them, showing
the results of the calculations for two representative examples, namely, for 
$4$ and $5$ atom wires. However the conclusions drawn from consideration of 
$4$ ($5$) atom wire remain valid for any even (odd) $N$ atom wire, provided the 
system is in ballistic regime. Moreover, we use the convention, in which the 
index $i$ labels the wire atoms, while $k$ indicates the position of the tip 
with respect to the wire atoms.

Let us first discuss the modifications of the wire density of states (DOS) due 
to the side-coupled impurity. The local DOS of $i$th atom in a wire is related 
to the corresponding diagonal element of the retarded GF, i.e. 
$\rho_i(\omega) = 
-\frac{1}{\pi} \sum_{\sigma} {\sf Im} G^r_{ii\sigma}(\omega)$. Similarly, for 
the impurity 
$\rho_a(\omega) = 
-\frac{1}{\pi} \sum_{\sigma} {\sf Im} G^r_{aa\sigma}(\omega)$. 

\begin{figure}[h]
 \begin{center}
  \resizebox{0.65\linewidth}{!}{
   \includegraphics{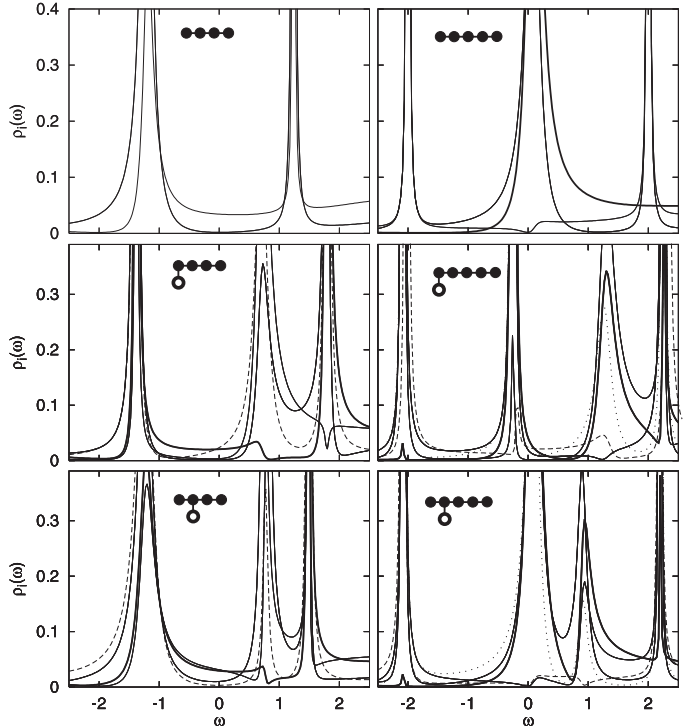}}
 \end{center}
 \caption{\label{Fig2} \footnotesize The local DOS $\rho_i(\omega)$ in various 
          wire-impurity configurations indicated in the insets. The left panels 
	  show $\rho_i(\omega)$ for $4$ atom and the right ones for $5$ atom 
	  wire, respectively. All the wire atomic energies as well as the tip
	  apex energy are assumed to be zero 
	  ($\varepsilon_w = \varepsilon_0 = 0$), while the impurity energy 
	  level is $\varepsilon_a = 1$. Other parameters are described in the 
	  text. Thin solid lines correspond to the DOS on first wire atom 
	  ($i = 1$ from left), thicker solid line is for $i = 2$, thickest 
	  solid line is for $i = 3$, $i = 4$ is described by dashed, and 
	  $i = 5$ by dotted line.}
\end{figure}

Figure \ref{Fig2} shows a local DOS of a wire consisted of $4$ (left panels) 
and $5$ atoms (right panels) in various impurity-wire configurations, indicated
in the insets to the panels.
The top panels show the wire local densities of states in the case when there 
is no impurity attached, i.e. for clean wire. In this case, for $N$ atom wire 
the local DOS on $i$th atom is the same as on $N+1-i$ atom 
($\rho_i(\omega) = \rho_{N+1-i}(\omega)$), i.e. there is a symmetry with 
respect to the middle of the wire, thus the only non-equivalent 
$\rho_i(\omega)$ are plotted. For $4$ atom wire the $\rho_i(\omega)$ has 
similar structure ($i = 1 = 4$ - thin solid line and $i=2 = 3$ - thicker solid 
line) featuring small DOS at the Fermi energy $E_F = 0$ and resonances 
corresponding to the wire atomic energies $\varepsilon_w = 0$ split by the 
hopping $t_w = 2$. On the other hand, for $5$ atom wire there are large DOS at
the $E_F$, and the resonances are eventually split by $t_w$. Such a behavior is
similar to the case of two terminal geometry in which the end wire atoms are 
connected to electrodes \cite{Muller,Smit,TK_2,Orellana,Liu,Zeng,MK_4}, i.e. a 
local minimum at the Fermi energy $E_F = 0$ for even number of atoms in a wire 
and local maximum at $E_F$ for odd atom wire and can be explained in terms of 
bonding, antibonding and nonbonding states \cite{Liu}. In case of even atom 
wire there are always bonding ($\omega < E_F$) and antibonding ($\omega > E_F$) 
states. When $N$ is odd, there is additional nonbonding state, which is 
situated at exactly the same position as the original atomic level 
($\omega = \varepsilon_w$), thus giving large DOS at $E_F$, as 
$\varepsilon_w = E_F$ in our case. Interestingly, for the second (or second 
from the end) atom in an odd atom wire (thicker solid line in the top right 
panel), the situation is quite different, namely, the local DOS has very small 
value at the $E_F$. It turns out that this is a general tendency in odd atom 
wires, i.e. every second atom in an odd atom wire shows small DOS at the Fermi 
energy. This could be understood in the following way. Let us forget for a 
moment about the STM tip, as we mentioned previously, that the STM tip does not 
influence the wire DOS, and assume that the wire is coupled to the surface 
only. If we calculated number of non-equivalent electron paths starting and 
ending in the surface and passing through a given atom, it turns out that if 
the number of odd paths (passing through odd number of atoms) $N_o$ is larger 
than the number of even paths (enclosing even number of atoms) $N_e$, then the 
local DOS has a resonance at $E_F$. Thus it behaves as in the case of odd atom 
wire in two terminal geometry with the end atoms coupled to the leads. For 
example, for the first atom in $5$ atom wire (thin solid line in the right top 
panel of Fig. \ref{Fig2}) we get $5$ non-equivalent odd paths and $4$ even 
paths, thus $N_o > N_e$ and the resonance at the $E_F$ is produced. Similarly, 
for the third atom, $N_o = 9$ and $N_e = 8$, and again $N_o > N_e$, thus we get 
the resonance at the Fermi energy (see thick solid line in the top left panel 
of Fig. \ref{Fig2}). Opposite is also true, namely, if $N_e$ is larger than 
$N_o$, the local DOS has similar behavior as in the case of even atom wire, 
featuring small $\rho_i(E_F)$. For example, $N_o = 7$ and $N_e = 8$ for the 
second atom in $5$ atom wire (thicker solid line on the same picture as 
previously). This seems to be true for any number of atoms in a wire in the 
present geometry. Of course, this is only intuitive picture, well working in 
the present case, namely, when the system is in ballistic regime, so the phase 
coherence length is longer than the wire length. In general, in the presence of 
interactions (electron-electron, electron-phonon or scattering on magnetic 
impurities) the phase coherence length will be suppressed, and this simple 
picture can be no longer valid. In this case one has to calculate the 
contributions from different electron paths.

The middle and the bottom panels of Fig. \ref{Fig2} show the local densities of
states $\rho_i(\omega)$ at different wire atoms in the case when the impurity
with a single atomic level $\varepsilon_a = 1$ is attached to the first and 
second wire atom, respectively. The presence of the impurity introduces
asymmetry in the DOS with respect to the middle of the wire, and the condition
$\rho_i(\omega) = \rho_{N+1-i}(\omega)$ does not hold anymore, as a result 
$\rho_i(\omega)$ will be different on each wire atom. In both wires the main 
modification of the DOS features additional resonance and thus splitting by 
$\omega = t_a$ of the resonance around $\omega = 1$ ($\omega = 2$) in the case 
of $4$ ($5$) atom wire. The low energy behavior of DOS is little affected, in 
particular for $4$ atom wire. Accidentally, it may a little bit shift the zero 
energy resonance, leaving small DOS at $E_F$, as it is seen in the middle right 
panel of Fig. \ref{Fig2}. This will be also reflected in the linear 
conductance, as we will see later.

\begin{figure}[h]
 \begin{center}
  \begin{minipage}[c]{0.35\linewidth}
   \resizebox{\linewidth}{!}{
    \includegraphics{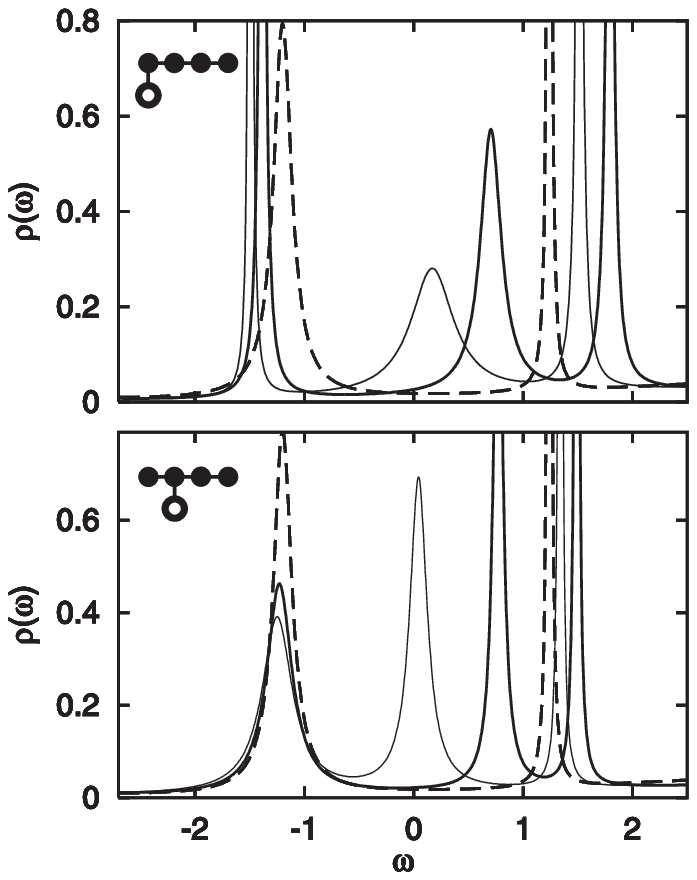}}
  \end{minipage}
  \hspace{0.08\linewidth}\begin{minipage}[c]{0.35\linewidth}
   \resizebox{\linewidth}{!}{
    \includegraphics{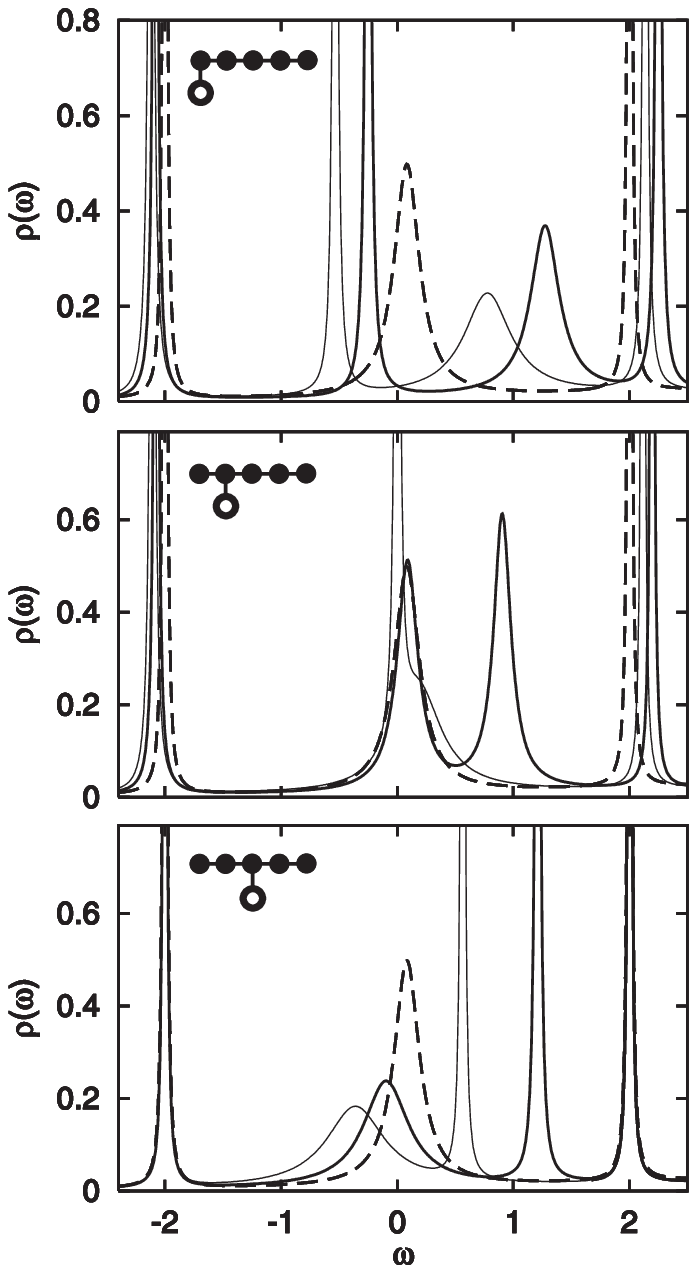}}
  \end{minipage}
 \end{center}
 \caption{\label{Fig3} \footnotesize Total density of states of a wire 
          consisted of $4$ atoms (left panels) and $5$ atoms (right panels) 
	  with side-coupled impurity. All the wire atomic energies 
	  $\varepsilon_w$ are equal to zero, while the impurity energy level 
	  $\varepsilon_a = 1$ (thick solid lines) and $\varepsilon_a = 0$ (thin
	  solid lines). The dashed lines represent the wire total DOS in the 
	  lack of the impurity. The insets show schematic configurations of the 
	  wire-impurity, i.e. the positions of the impurity with respect to the 
	  wire atoms.}
\end{figure}

In two terminal geometry, when the end wire atoms are connected to two 
different leads, the useful quantity is the total density of states, as in 
this case the conductance of the system is correlated with it \cite{TK_2}. On 
the other hand, in the STM geometry we do not expect that the behavior of the 
conductance will be governed by this quantity. To see this, we plotted the 
total (wire plus impurity) density of states $\rho(\omega)$ of a chain 
consisted of $4$ atoms (left panel) and $5$ atoms (right panel of Fig. 
\ref{Fig3}).
The total DOS is defined as a $\rho(\omega) = \sum_i \rho_i(\omega)/N$ for the 
wire with no impurity attached, and as a
$\rho(\omega) = (\sum_i \rho_i(\omega) + \rho_a(\omega))/(N+1)$ in the case 
when the impurity is present. We have distinguished two cases, namely, when the 
impurity has the same value of the atomic energy $\varepsilon_a = 0$ as for the 
wire atoms, and when it has different energy $\varepsilon_a = 1$. The former 
case would correspond to the situation when the wire atoms and the impurity are 
composed of the same material, while in the later one, the impurity is a 
different atom. 

First of all, if there is no impurity (dashed lines) the total DOS shows 
similar behavior as in the case of the end wire atoms coupled to the leads 
only. However one can notice different heights of particular resonances. This 
reflects an effect, known from the studies of two impurities on a surface, and 
associated with different (even-odd) symmetries of electron states in resulting 
system \cite{Newns}. When the impurity is introduced to the system, the total 
DOS is strongly modified due to the parameter $t_a$, which is responsible for 
the hoping between those subsystems. In this case $t_a$ is only two times 
smaller than the wire hoping $t_w$, thus one should expect modifications of the 
wire DOS. The impurity usually introduces additional resonance to the total 
DOS. In both cases the position of the resonance is around its atomic energy
(compare thin and thick solid lines in in the left panels of Fig. \ref{Fig3}), 
and slightly modifies the resonances coming from the wire atoms. This behavior 
only slightly depends on which wire atom is in close connection with the 
impurity. 

\begin{figure}[h]
 \begin{center}
  \resizebox{0.7\linewidth}{!}{
   \includegraphics{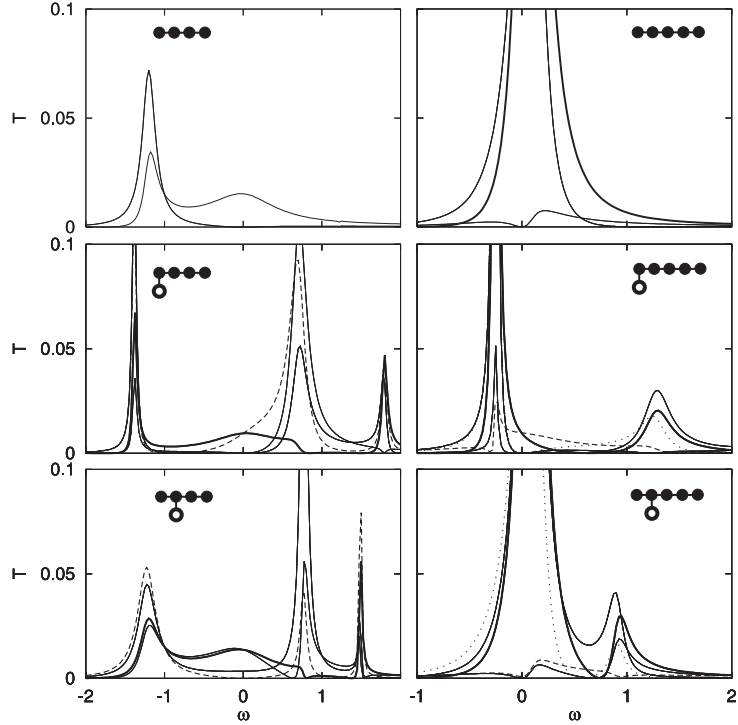}}
 \end{center}
 \caption{\label{Fig4} \footnotesize Transmittance $T(\omega)$ of the system in
          various wire-impurity configurations indicated in the insets and
	  various positions of the STM tip. The left panels show $T(\omega)$ 
	  for $4$ atom and right panels for $5$ atom wire, respectively. The 
	  impurity energy level is now $\varepsilon_a = 1$. Thin solid line
	  corresponds to the STM tip above first wire atom ($k = 1$ from left), 
	  thicker solid line is for $k = 2$, thickest solid line is for 
	  $k = 3$, $k = 4$ is described by dashed, and $k = 5$ by dotted line.}
\end{figure}

Let us now turn to the transmittance $T(\omega)$, defined by Eq. 
(\ref{transmit}). Figure \ref{Fig4} shows the transmittance of the system in
various STM tip-wire-impurity configurations for $4$ atom (left panels) and $5$
atom wire (right panels).
While the transmittance depends on the STM tip position now, it is possible to 
identify the impurity induced contribution to $T(\omega)$. The impurity
introduces additional resonance to $T(\omega)$ around its atomic energy, i.e. 
at $\omega = \varepsilon_a$ (compare the left panels of Fig. \ref{Fig4}). In 
both cases, i.e. for $4$ and $5$ atom wires, the transmittance is correlated 
with the local density of states, and the presence of the impurity leads to 
similar modifications, especially for $\omega = 0$ (compare Fig. \ref{Fig2}), 
although the transmittance is now more asymmetrical. This is in contradiction 
with the transport along the wire, as in that case the transmittance depends on 
the total DOS. Here it depends on the local DOS. Explanation is simple and 
intuitive. In the transport along the wire, all the local DOS equally 
contribute to the transmittance \cite{TK_2}. In the present case, the transport 
takes place mainly through the wire atom, just below the STM tip, thus 
$T(\omega)$ depends mainly on the local DOS of a particular atom.

The presence of the impurity usually leads to higher values of $T(\omega)$ for 
the resonances of the wire origin. Such a behavior is in contradiction with the 
behavior of $T(\omega)$ when the transport takes place along the wire. In the 
former geometry this leads to the reduction of the conductance 
\cite{Kang}-\cite{Stefanski}. Here however, the situation is different, as the 
impurity is also connected to the same lead as the wire is, thus the impurity 
provides additional tunneling channel. Nevertheless, the zero energy
transmittance is influenced by the impurity, similarly as the local DOS 
(compare Fig. \ref{Fig2}). 

At this point we would like to comment on the Fano effect, as there are many 
channels for electron tunneling from STM tip to the surface. Even the tip is
'coupled' to a single wire atom, the electron can leave the wire entering the 
surface electrode through different wire atoms. This is particularly well 
visible for STM tip placed above second atom in odd atom wire even without 
impurity attached (thicker solid line in top right panel of Fig. \ref{Fig4}). 
In this case odd number atoms in the wire have large density of states at the 
Fermi energy, while even atoms have small DOS at $E_F$ (see Fig. \ref{Fig2}). 
When the STM tip is placed above first (thin solid line) or third (thickest 
solid line) an electron can tunnel through undelaying atom directly to the 
surface because neighboring atoms have small DOS at $E_F$. The Fano effect is 
negligible in this case. On the countrary, when the tip is placed above second 
atom, which has small DOS at $E_F$ (thicker solid line), the electron has three 
tunneling channels to the surface, due to large values of DOS at neighboring 
atoms. It can directly tunnel to the surface or go through first or third wire 
atom. In this case the Fano effect is enhanced and is visible as a dip at the
Fermi energy. The presense of impurity slightly modifies the above picture, 
usually enhancing the Fano effect when the tip is above the wire atom with 
attached impurity. Compare thin solid line in midle right panel of Fig.
\ref{Fig4} and thicker solid line in the bottom right panel. We do not observe
Fano effect for even atom wire because DOS at $E_F$ is always small (see Fig.
\ref{Fig2}).

\begin{figure}[h]
 \begin{center}
  \resizebox{0.7\linewidth}{!}{
   \includegraphics{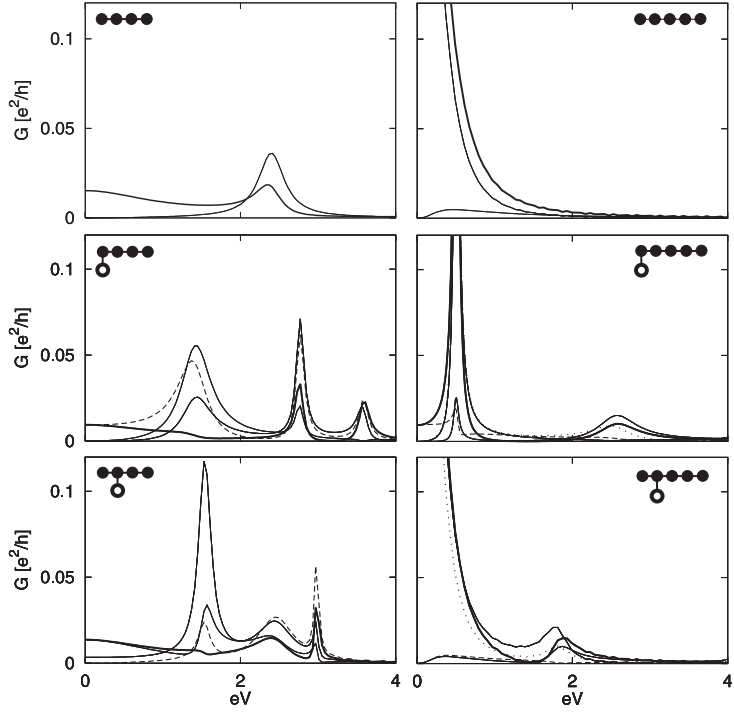}}
 \end{center}
 \caption{\label{Fig5} \footnotesize Corresponding differential conductance vs. 
          bias voltage of the system in the same impurity-wire configurations,
	  as in Fig. \ref{Fig4}. The parameters and the meaning of the curves
	  are the same as those in Fig. \ref{Fig4}.}
\end{figure}

Similar effects are reflected in differential conductance $G = dI/d(eV)$, shown 
in Fig. \ref{Fig5}. 
Again, the presence of the impurity leads to a increase of $G$ around 
$eV = 2 \varepsilon_a$ (note that $eV = \mu_s - \mu_t$ and $\mu_s = - \mu_t$), 
thus can help us in studying of the properties of impurities attached to wire 
in real STM experiments. It also modifies the structures coming from the wire 
atoms, also leading to larger values of $G$. Note however, that the values of 
$G$ are much smaller ($0.12 \; e^2/h$ for $4$ atom and $0.55$ for $5$ atom 
wire) than the conductance unit, i.e. $e^2/h$ per tunneling channel. This is 
due to small value of the STM tunneling parameter $t_0$. For $t_0 = 1$, the 
conductance reaches the unitary limit, i.e. $2 e^2/h$ in our case (two spin 
channels). On the other hand, the linear conductance, i.e. $G(eV = 0)$ 
strictly follows the local density of states, similarly as the zero energy 
transmittance (compare Fig. \ref{Fig5} and Figs. \ref{Fig2} and \ref{Fig4}). 
In a special case when the impurity is coupled to the first atom in $5$ atom 
wire, the zero bias maximum is split and the linear transport is strongly 
reduced for any wire atom (compare the middle right panel of Fig. \ref{Fig2}). 

\begin{figure}[h]
 \begin{center}
   \resizebox{0.47\linewidth}{!}{
    \includegraphics{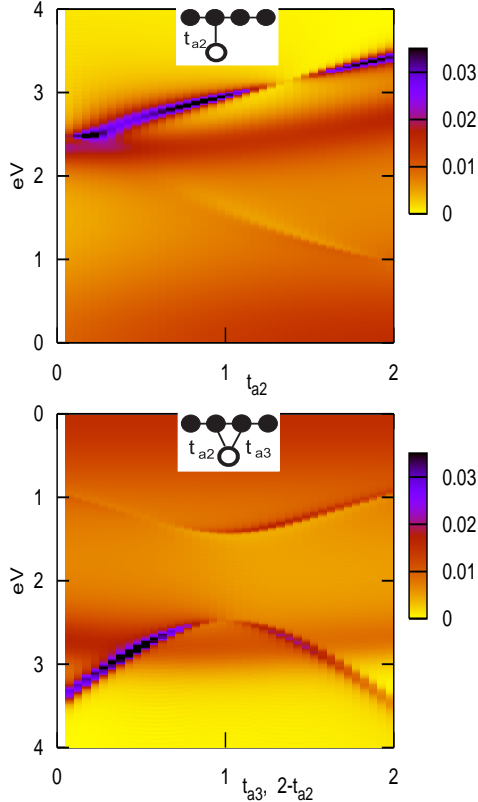}}
 \end{center}
 \caption{\label{Fig6} \footnotesize Differential conductance vs eV and $t_a$.
          Top panel represents the situation when the impurity is coupled to 
	  the second wire atom via the parameter $t_{a2}$. The bottom panel 
	  shows $G$ in the case when the impurity is coupled to the second and 
	  the third wire atom via $t_{a2}$ and $t_{a3}$ ($t_{a3} = 2 -t_{a2}$), 
	  respectively.}
\end{figure}

So far we have discussed the various configurations, in which impurity was
coupled to a single atom wire. This would correspond to a situation in which 
the impurity is placed close to one particular wire atom. Now, the question 
arises what will happen if the impurity is side-coupled to two wire atoms, 
i.e. is placed aside between two wire atoms. Figure \ref{Fig6} shows the 
conductance changes due to different couplings of the impurity to one atom (top 
panel) and two atoms in a wire (bottom panel).
Both panels show the conductance maps vs. bias voltage $eV$ and corresponding
couplings $t_{a2}$ ($t_{a3}$) (see the insets to the figure). In first 
configuration (top panel) the impurity is coupled to one wire atom, and the 
differential conductance has a maximum in the plane ($eV,t_{a2}$). The position
of the maximum changes linearly with $eV$ and $t_{a2}$. On the other hand, in 
second configuration, when the strength of the impurity coupling changes 
(bottom panel), i.e. for various positions of the impurity between second and 
third wire atom, we see that the largest values of $G$ are obtained when the 
impurity is close to the second atom ($t_{a3} \approx 0$, $t_{a2} \approx 2$) 
and slightly smaller close to the third wire atom 
($t_{a2} \approx 0$, $t_{a3} \approx 2$). For intermediate values of $t_{a2}$
($t_{a3}$) the conductance is strongly reduced, in particular, when 
$t_{a2} = t_{a3}$. This is due to the Fano effect, which is further enhanced in 
this case, and leads to similar reduction of $G$ in the case of two terminal 
geometry with the end wire atoms coupled to the leads 
\cite{Kang}-\cite{Stefanski}. The fact, that the maximal values of the 
differential conductance show asymmetry with respect to $t_{a2} = t_{a3}$ 
point, i.e. they are different for the impurity near the second wire atom and 
near the third wire atom steams from the fact that STM tip is placed above the 
second wire atom.

Finally we would like to comment on the Coulomb interactions, as the present
model completely neglects them. We expect qualitative modifications of the 
results (like even-odd oscillations of density of states), especially at low 
temperatures where the Coulomb blockade and the Kondo effect take place. The 
picture can change drastically \cite{MK_1}. The tunneling can be enhanced when  
the wire energy levels are placed below the Fermi energy and the wire is
strongly coupled to the surface (Kondo effect) or suppressed when it is weakly 
coupled, leading to the Coulomb blockade. The presense of additional tunneling
channel (due to the Kondo effect) can also enhance the Fano effect. On the 
other hand, we do not expect qualitative modifications in the mixed valence and 
the empty regimes, i.e. when the wire energy levels are above the Fermi energy.
The present model can be succesfully applied to study monoatomic (Au, Ag, Pb)
chains grown on various vicinal surfaces \cite{Himpsel}-\cite{MK_2}.

%%%%%%%%%%%%%%%%%%%%%%%%%%%%%%%%%%%%%%%%%%%%%%%%%%%%%%%%%%%%%%%%%%%%%%%%%%   
\section{Conclusions}
\label{conclusions}
%%%%%%%%%%%%%%%%%%%%%%%%%%%%%%%%%%%%%%%%%%%%%%%%%%%%%%%%%%%%%%%%%%%%%%%%%%   

In conclusion we have studied STM tunneling through a quantum wire with a
side-coupled impurity in various coupling configurations, i.e. the impurity was
coupled to various atoms in a wire, and moreover, it was also connected to more
than single wire atom. We have found that the impurity strongly modifies the
transport properties. In particular it always produces a resonance around its
atomic energy, which can be seen in differential conductance vs. bias voltage.
Moreover, we have also shown that if the impurity is side-coupled to two wire
atoms with equal strength, it leads to the suppression of the conductance, 
which is a hallmark of the Fano effect in such a system. Those studies could be
potentially useful in STM studying of the impurity induced modifications of 
the wire properties.

%%%%%%%%%%%%%%%%%%%%%%%%%%%%%%%%%%%%%%%%%%%%%%%%%%%%%%%%%%%%%%%%%%%%%%%%%%%%%%%

\section*{Acknowledgements}

This work has been supported by the Grant No N N202 1468 33 of the Polish 
Ministry of Education and Science.

%%%%%%%%%%%%%%%%%%%%%%%%%%%%%%%%%%%%%%%%%%%%%%%%%%%%%%%%%%%%%%%%%%%%%%%%%%%%%%

%%%%%%%%%%%%%%%%%%%%%%%%%%%%%%%%%%%%%%%%%%%%%%%%%%%%%%%%%%%%%%%%%%%%%%%%%%   
\section*{Appendix}
\label{apendix}
%%%%%%%%%%%%%%%%%%%%%%%%%%%%%%%%%%%%%%%%%%%%%%%%%%%%%%%%%%%%%%%%%%%%%%%%%%   

In this appendix  we give analytical solutions/relations for the local density 
of states (LDOS) of a wire disturbed by an adatom. LDOS is connected with the 
retarded Green's function by the following relation
\begin{eqnarray}
LDOS_i(\omega)={-1\over \pi} Im G^r_{ii}(E)
\end{eqnarray}
where $G^r(E)$ can be obtained from the equation of motion for Green functions.
 Using Eq. 2 one can write
\begin{eqnarray}
\hat G^r_{ii}(\omega)=(\omega \hat 1 - \hat H)^{-1} = 
\hat Z^{-1}_{ii}=\frac{{\texttt{cof}}{\hat Z_{ii}}}{\det{\hat Z}}
\end{eqnarray}
where $\texttt{cof}{\hat Z_{i}}$ is the algebraic complement of the matrix 
$\hat Z$ (cofactor). In our calculations we assume the same coupling strengths 
between atoms, $t_w$, and the same single particle energies $\varepsilon_w$ for 
all atoms in the wire. It is worth noting that the STM is weakly coupled with 
the wire and thus it does not affect the wire density of states (in Hamiltonian 
Eq. (\ref{Hamilt}) we put $t_0=0$).

First we consider the case of a linear wire without an adatom i.e. $t_a=0$. In 
this case LDOS can be obtained from the relation 
$LDOS_{ii}(\omega)=-Im {\texttt{cof}{\hat A^N_{ii}} / \pi \det{\hat A^N}}$ 
where
\begin{eqnarray}
\hat A^N=\hat A^N_{k_1k_2}=
(\varepsilon-\varepsilon_w)\delta_{k_1,k_2}-t_w(\delta_{k_1,k_2+1}+
\delta_{k_1+1,k_2})-\Sigma_S
\end{eqnarray}
and $\Sigma_S=-i\Gamma/2$, $k_1, k_2 = 1, ..., N$. To obtain LDOS one needs to 
know the determinant of $\hat A^N$ which can be expressed as follows
\begin{eqnarray}
\det \hat A^N=\det\hat A^N_0-\Sigma_S \sum_{k=1}^N \det\hat A_0^{N-k} 
\sum_{l=1}^{k-1}(-t_w)^{k-l} \det\hat A_0^{l-1}
\nonumber\\ 
-\Sigma_S \sum_{k=1}^N \det\hat A_0^{k-1} \sum_{l=k}^{N} (-t_w)^{l-k} 
\det\hat A_0^{N-l}
\end{eqnarray}
The matrix $\hat A^N_0$ corresponds to an isolated wire (non-coupled with the 
surface) and is a tridiagonal one, $\hat A^N_0=\hat A^N_{k_1k_2}+\Sigma_S$, and 
can be expressed analytically in terms of Chebyshev polynomials of the second 
kind \cite{TK_2}.

The LDOS for a wire coupled with an adatom can be obtained from the relation 
$LDOS_{ii}(\omega)=-Im { \texttt{cof}{\hat B^{N+1}_{ii}} / 
\pi \det{\hat B^{N+1}}}$ where
\begin{eqnarray}
\hat{B}=\left(%
\begin{array}{cc}
  \hat A^N & \hat X^T \\
  \hat {X} & \varepsilon-\varepsilon_a \\
\end{array}%
\right)
\end{eqnarray}
and  $\hat X$ is a vector describing the couplings adatom-QW and 
adatom-surface, $\hat{X}=-(\Sigma_S, ... ,t_a+\Sigma_S,...,\Sigma_S)$ ($j$-th 
atom of a wire is connected with the adatom. After some algebra the determinant 
of the above matrix can be written in the form
\begin{eqnarray}
\det \hat B^{N+1}=(\omega-\varepsilon_a) \det\hat A^N-\Sigma_S 
\det\hat A^N_0+t_a \texttt{cof}{\hat A^{N-1}_{jj}} + \nonumber\\
 2t_a \sum_{k=1}^N (-1)^{k+j} (t_w) ^{|j-k|} \det\hat A_0^{\min(k,j)-1}  
 \det\hat A_0^{N-\max(k,j)}
\end{eqnarray}
Similar equation one can write for $\texttt{cof}{\hat A^{N}}$ and 
$\texttt{cof}{\hat B^{N+1}}$ which also can be expressed in terms of 
$\det \hat A_0$ but have more complicated structure. For the case $\Sigma_S=0$ 
(there is no surface under the wire or the coupling wire-surface is very weak) 
one can easily find
\begin{eqnarray}
\det \hat B^{N+1}&=&(\omega-\varepsilon_a) \det\hat A^N_0-t_a^2 
\det\hat A^{j-1}_0 \det\hat A^{N-j}_0 \\
\texttt{cof}{\hat B^{N+1}_{ii}}&=&(\omega-\varepsilon_a) \det\hat A^{j-1}_0 
\det\hat A^{N-j}_0+t_a^2 \det\hat A^{\min(i,j)}_0
\det\hat A^{|i-j|-1}_0 \det\hat A^{N-\max(i,j)}_0
\end{eqnarray}
and the LDOS can be obtained fully analytically. It is worth noting that the 
determinants $\det \hat A$, $\det \hat B$ and $\det \texttt{cof} \hat B$
are obtained for arbitrary $j$ (connection adatom-QW atom) and $N$ (even or 
odd). The minima of $|\det \hat B|^2$ determine high value of LDOS and the 
conductance through the system.

%%%%%%%%%%%%%%%%%%%%%%%%%%%%%%%%%%%%%%%%%%%%%%%%%%%%%%%%%%%%%%%%%%%%%%%%%%%%%%

%%%%%%%%%%%%%%%%%%%%%%%%%%%%%%%%%%%%%%%%%%%%%%%%%%%%%%%%%%%%%%%%%%%%%%%%%%   

\end{document}